\documentclass[aps,prb,twocolumn,showpacs]{revtex4}
\usepackage{amsmath}
\usepackage{epsfig}
\usepackage{graphics}
\usepackage{hyperref}

 \renewcommand{\Im}{{\rm Im}}

\newcommand{\bari}{{\bar \imath}}

\def \o{\omega}    
    
\def \e{\epsilon}
   \def \d{\delta}

\def \H{\hat{\cal H}}
\def \half{{1 \over 2}}
\def \dag{\dagger}

\def \ra{\rightarrow}

\def \be{\begin{equation}}
\def \ee{\end{equation}}
\def \bea{\begin{eqnarray}}
\def \eea{\end{eqnarray}}
\def \bea*{\begin{eqnarray*}}   
\def \eea*{\end{eqnarray*}}
\def \v{\mathbf}

\def\r{\right}
\def\l{\left}

\begin{document}

\title{Charge oscillations in Quantum Dots: Renormalization group and Hartree method calculations}

\author{Michael Sindel,$^{1}$ Alessandro Silva,$^{2,3}$ Yuval Oreg,$^{3}$ and Jan von Delft$^{1}$}
\email[ Correspondence to ]{michael.sindel{@}physik.lmu.de}
 \affiliation{
 $^1$Physics Department and Center for NanoScience, LMU M\"unchen, 80333 M\"unchen, Germany\\
$^2$Center for Materials Theory, Department of Physics and Astronomy, Rutgers University, Piscataway, NJ 08854, \hspace{-2.2mm} USA\\
$^3$Department of Condensed Matter Physics, The Weizmann Institute
of Science, Rehovot 76100, Israel}

\date{\today}

\begin{abstract}

We  analyze  the local level occupation of a spinless, interacting
two-level  quantum  dot  coupled to two leads by means of Wilson's
numerical  renormalization  group  method.  A  gate voltage sweep,
causing  a  rearrangement  of  the  charge  such that the system's
energy  is  minimized,  leads to oscillations,  and sometimes even
inversions,   in   the  level  occupations. We find that these
oscillations, qualitatively understandable by a simple Hartree
analysis,  are generic and occur in a wide range of system
parameters. By  allowing  a  relative sign in one tunneling matrix
element  between  dot and leads, we extend our findings to more
generic models. Experimental applications and the qualitative effect
of spin are discussed.
\end{abstract}


\maketitle
\marginparwidth=10cm
 \section{Introduction}
 The Coulomb blockade (CB)~\cite{CB} is one of the most basic and
 fundamental phenomena in quantum dot (QD) physics: according to the
 standard single-electron tunneling picture for transport through
 small QD's, electrons can only pass through the dot one by one if the
 Coulomb charging energy $U$ is the dominant energy scale.  Likewise,
 if the gate voltage (applied to a plunger gate near the QD) is
 changed, resulting in a rigid shift of the single-particle spectrum
 of the QD relative to the Fermi energy of the leads, the
 single-particle levels of the QD are naively expected to be filled
 ``one by one''. Many basic single-electron devices and QD-based qubit
 proposals~\cite{Loss} are based on this simple picture.

In this paper we point out that under quite generic conditions, the
charging process is actually more complicated,
and the occupation functions for the single-particle levels can show
some rather complex, non-monotonic behavior as a function of gate
voltage, deviating considerably from the standard CB picture [``one by
one'' filling]. This complex behavior turns out to be rather generic
for a QD coupled to leads that can supply the QD with electrons, and
originates from the competition between the QD-leads coupling
$\Gamma$, and the intrinsic energy scales of the QD, namely its charging energy
$U$ and levels spacing $\Delta$.

In order to study this competition, it suffices to consider a very
simple model: a spinless interacting two-level Anderson model
(2LAM), consisting of a lower and an upper QD level ($\e_{\ell}$,
$\e_u$) with level spacing\cite{ADeltaU} $\Delta\equiv
\e_u-\e_{\ell}$. Analyzing this model using both the numerical
renormalization group (NRG) \cite{Wilson} and a self-consistent
Hartree approximation, we study in detail the evolution of the
occupation of the single particle levels as a function of gate
voltage at $T=0$ for various values of the QD parameters ($\Gamma$,
$\Delta$, $U$).  The {\it generic picture} of the charging process
emerging from this analysis is the following: for any finite
coupling to the leads $\Gamma$ ($\not \ll \Delta$), sweeping the
levels towards the Fermi level of the leads (by tuning the gate
voltage) causes the occupations of both the lower {\it and} the
upper level to {\it increase at
  comparable rates}.  This process continues until one level takes
over and becomes more occupied than the other. At this point the
electron that occupies this level electrostatically repels the other
level, pushing up its energy and thereby emptying it.  As a result the
occupation of the other level {\it performs an oscillation} as the
gate voltage is swept.  The naive QD charging scheme, in which every
step of the CB staircase is associated with the filling of {\it only
  one} single particle level in the QD, is only achieved when these
charge oscillations are small, i.e. for $\Gamma \ll \Delta$.  Below,
we discuss in detail the physics of ``charge oscillations'' and the
dependence of their amplitude and form on the system parameters.  In
particular, we discuss under which conditions the amplitude of these
oscillations can be made so large as to cause a significant
``occupation inversion'', i.e.\ a situation for which the occupation
of the lower level is {\it smaller} than that of the upper level. It
is important to mention that more general models, e.g.\ including
spin, show effects similar to those described above.

The rest of the paper is organized as follows. In Section II we
describe in detail the model we consider. In Section III, we present
the result of the NRG calculation for the evolution of the occupation
of the two levels as a function of gate voltage [Section III A], and a
simple and qualitative understanding of the emerging picture in terms
of an Hartree approach [Section III B]. In Section IV, we present a
detailed analysis of the phenomenon of occupation inversions, by
studying the 2LAM in the case where the two levels are unequally
coupled to the leads. Finally, in Section V we summarize our
conclusions, and discuss the robustness of the effects obtained when
including spin, and possible experimental tests for our predictions.

\section{Model}

 We consider a spinless 2LAM with Hamiltonian
$\H=\H_d+\H_l+\H_{ld}$. (The spinful case will be considered
briefly in Section~\ref{sec:summary}.) Two leads, identical,
noninteracting and in equilibrium, are described by
$\H_l=\sum_{ka}\e_{ka}c^{\dag}_{k a}c_{k a}$, where $c_{k
a}^{\dagger}$ creates  an electron with energy $\e_{k}$ in lead
$a=L,R$. The isolated QD \mbox{is described by}
\begin{eqnarray} \label{Ham_Dot}
\H_d&=&\sum_{i= u, \ell}\e_{i} d^{\dag}_{i}d_{i}^{\phantom{\dag}} +U
\hat{n}_{\ell}\hat{n}_{u},
\end{eqnarray}
where $d_{i}^{\dagger}$ creates an electron in the QD in level
$\e_{i}$ ($i=\ell,u$), measured w.r.t.\ the Fermi level defined by the leads,
$\hat{n}_{i}=d^{\dag}_{i}d_{i}^{\phantom{\dag}}$ is the number
operator,  and $U$ is the charging energy, which we fix at
$U=0.2 D$   throughout this paper, $2 D$ being the bandwidth.
Finally, the tunneling between the QD and the leads is described
by $\H_{ld}=\sum_{kia} (V^{\phantom{\dag}}_{kia} c^{\dag}_{k a}
d^{\phantom{\dag}}_{i} +h.c.)$.
We consider $k$-independent tunneling matrix elements
$V_{kia}=V_{ia}$, which are $L$-$R$-symmetric in
magnitude~\cite{validity}, $V_{uL}=V_{uR}=V_{u}$ and $V_{\ell L}=
s V_{\ell R} = V_{\ell}$, but with a possible relative phase
$s=\exp(i\phi)$ between the $L$ and $R$ matrix elements of the
lower level~\cite{Silva,Entin,Sun,Boese}.  Time reversal symmetry implies
$\phi=0,\pi$ (hence $s=\pm 1$).  The corresponding bare level
widths are $\Gamma_i=2\pi\rho V_i^2$, where $\rho$ is the density
of states in the leads.

\begin{figure}[t]
\centering
\includegraphics[width=1.0\columnwidth]{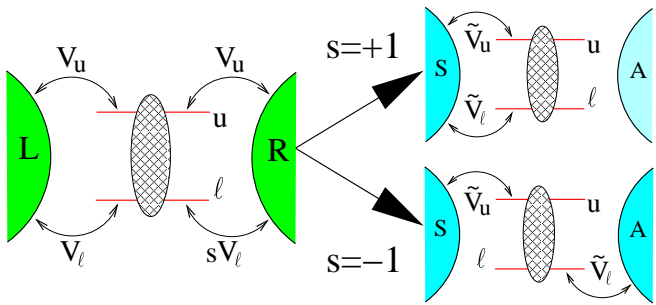}
\caption{Schematic depiction of the model. The sign $s$ between the
tunneling amplitudes $V_{\ell R}$ and $V_{\ell L}$ determines
whether the two dot levels couple to the {\em same} (symmetric)
channel ($s=+1$, upper mapping), or to two {\em different} channels
($s=-1$, lower mapping), with strength ${\tilde V}_i=\sqrt{2}V_i$,
$i=\ell,u$.} \label{fig:model}
\end{figure}

The two possible choices for $s=\pm 1$ lead to two distinct models
(see Fig.~\ref{fig:model}): (i) for $s=+1$, both local levels
couple to the same channel, namely the symmetric linear combination of
the left and the right lead $\l(c_{k L}+ c_{k R}\r)$,  while the
  antisymmetric combination $\l(c_{k L}- c_{k R}\r)$ decouples
  completely; (ii) for $s=-1$, the upper and lower local levels
  couple to different channels, namely to the symmetric or
  antisymmetric combinations, 
  respectively~\cite{FMleads}.

We shall denote the ground state
  expectation value of the occupation of level $i$ by $n_i=\langle\hat{n}_i\rangle$, and
  parameterize the gate voltage by the average bare level position
  $\epsilon\equiv\l(\e_{\ell}+\e_u\r)/2$. For
  $\Gamma_{\ell}=\Gamma_{u} = \Gamma$, this parameterization reveals
particle-hole symmetry~\cite{phsymmetry} around $\e^*=-U/2$, namely
$n_u(\e+\e^*)=1- n_{\ell}(\e^*-\e)$, independent of $\Delta$ and
$\Gamma$.  If $\Gamma_{\ell}\neq\Gamma_{u}$, this symmetry is broken
for both $s=\pm 1$.

\section{Charge oscillations}

In this section, we study the evolution of $n_i$ ($i=1,2$) as a
function of the gate voltage $\epsilon$, and discuss the physical
origin of non-monotonic occupation of the two levels. We shall first
present the results of our NRG \cite{Wilson} calculations for $n_i$,
and then use a simple Hartree analysis to gain some insight into the
NRG results.

\subsection{NRG calculations }

\begin{figure}[h]
\centering \includegraphics[width=1.0\columnwidth]{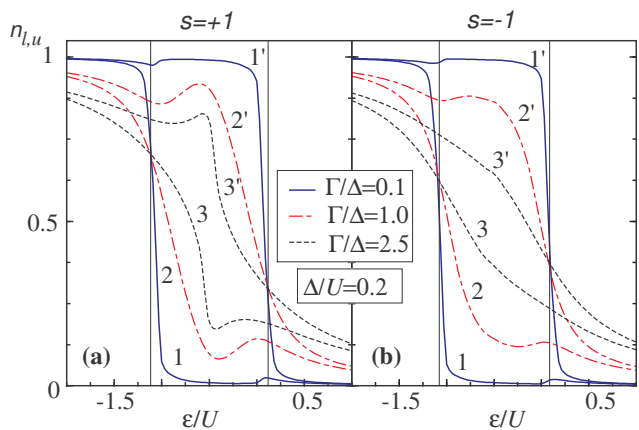}
\caption{NRG results for the occupation of the lower $[n_{\ell}$,
(curves
  $1'$ - $3'$)$]$ and the upper level $[n_u$, (curves $1$ - $3$)$]$
   for fixed $\Delta$ and different values of $\Gamma$,
for (a): $s=+1$ and (b): $s=-1$.
The dotted lines indicate where the lower/upper level
crosses the Fermi level, at $\e=\Delta/2$ and $\e=-U-\Delta/2$.}
\label{fig:varyGamma}
\end{figure}

We start our analysis by considering equally coupled levels,
$\Gamma_{\ell}=\Gamma_{u}=\Gamma$, and use the NRG~\cite{Wilson} to
calculate the $\e$-dependence of $n_i$. Naively one may expect that
if the QD is initially empty, the QD levels get occupied
monotonically one by one as $\e$ is decreased, the usual CB
behavior.  In other words, first the occupation of the lower level
would be expected to increase monotonically as $\epsilon_{\ell}$
crosses the Fermi level, and subsequently $n_u$ would increase as
$\epsilon_u+U$ approaches it.  However, our NRG results
[Fig.~\ref{fig:varyGamma}] show that this intuitive picture is valid
only if the coupling to the leads is much smaller than the dot level
spacing, $\Gamma\ll \Delta$.  In particular, when $\Gamma \gtrsim
\Delta$, $n_{\ell}$ and $n_u$ show a {\it
  non-monotonic} $\e$-dependence, characterized by
\it charge oscillations \rm of $n_u$ (or $n_{\ell}$) when the lower
(or upper) level crosses the Fermi level~\cite{Berkovits03}.  The oscillation in $n_u$
occurs because as soon as the lower level begins to be occupied
significantly, the system can gain charging energy by additionally
filling the lower level and emptying the upper level (an analogous
argument works for $n_{\ell}$).

To explore how strongly these charge oscillations vary with $\Gamma$,
$\Delta$ and how they are affected by the sign of $s$, we show in
Fig.~\ref{fig:varyGamma} the behavior of $n_i(\e)$ for variable
$\Gamma$ and $s=\pm 1$, keeping $\Delta$ fixed (at $0.2 U$).  In the limit
$\Gamma/\Delta\ll 1$
level $\ell$ becomes occupied rather suddenly (curve $1'$) when it crosses
the Fermi level at $\e=\Delta/2$,
and similarly for the upper level at $\e=-U-\Delta/2$ (curve $1$).
In addition to this typical CB behavior, we observe, even for the
smallest $\Gamma$ considered ($\Gamma=0.1 \Delta$), a tiny
non-monotonicity or {\it charge oscillation} in $n_u$ ($n_{\ell}$)
roughly at that $\e$ where the occupation of the lower (upper) level
increases sharply from $0$ to $1$.  A gradual increase of $\Gamma$
towards $\Gamma/\Delta\approx 1$ results in a strengthening of these
charge oscillations.  In the limit $\Gamma/\Delta\gg 1$,
Fig.~\ref{fig:comp_HF_NRG}(c) and (d), the monotonic increase in the
occupation is recovered, though the actual dependence of $n_i$ on $\e$
depends strongly on the sign of $s$.

\subsection{Hartree approach}

 A simple and qualitative understanding of the
charge oscillations observed in our NRG results can be obtained in
the framework of a self-consistent Hartree approximation (scHA), by
studying the evolution of the Hartree levels as function of
$\epsilon$. This scheme  accounts for the interaction by replacing
the bare levels $\e_i$ of the {\em noninteracting} problem, with the
corresponding Hartree levels
\begin{equation} \label{Hartree_level}
 \e_{i}\ra\e_i^H=\e_{i}+U n_{\bari},
\end{equation}
where ${\bari}=\ell/u$ if $i=u/\ell$. By integrating out the
leads, we obtain the effective noninteracting dot Hamiltonian
\begin{equation}
{\cal H}^{\rm eff}_d =
\l(\begin{array}{cc} \e^H_u-i\Gamma_u&\l(-i\sqrt{\Gamma_u\Gamma_{\ell}}\r)\d_{s,+1}\\
\l(-i\sqrt{\Gamma_u\Gamma_{\ell}}\r) \d_{s,+1}
&\e^H_{\ell}-i\Gamma_{\ell}\end{array} \r).
\end{equation}
The corresponding retarded dot Green's function, defined as
${\cal G}_{i \! j}^R(t)\equiv -i\theta(t)
\langle\{d_i(t),d_{j}^{\dag}(0)\}\rangle$, can be obtained exactly for
both values of $s$, by solving the matrix equation $\v{{\cal G}}^R
(\o)=(\o-\v{{\cal H}}_{{\rm eff }})^{-1}$.  To finally obtain the
Hartree approximation for the {\em interacting} Green's function, one
has to self-consistently calculate the average level occupation
$n_i(\e)$, using the $T=0$ relation
\begin{equation} \label{occupation}
n_{i}(\e)=-\frac{1}{\pi}\int_{-\infty}^{0}d\omega\; \Im \, {\cal
  G}^{R}_{ii}(\o,\e).
\end{equation}

\begin{figure}[b]
\centering \includegraphics[width=1.0\columnwidth]{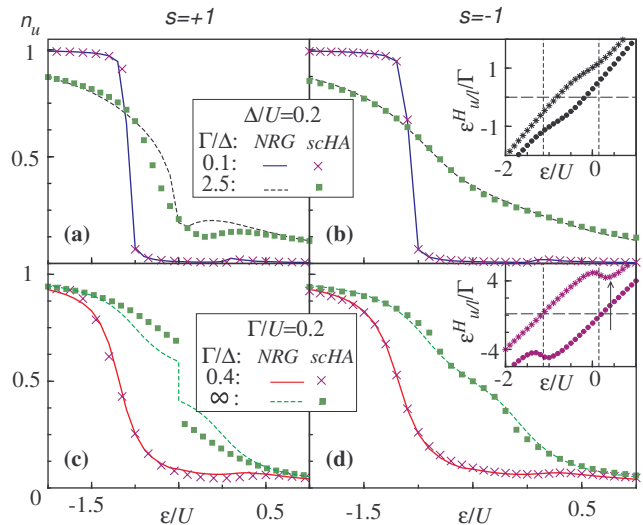}
\caption{Comparison of NRG and scHA results for $n_u(\e)$, for fixed
$\Delta$
  and variable $\Gamma$ [(a), (b)] or fixed $\Gamma$ and
  variable $\Delta$ [(c), (d)].  The naive scHA used for $s=+1$ works
  well for $\Gamma/\Delta<1$.  Insets in (b) and (d):
  Hartree levels $\e^H_{\ell}$ (circles) and $\e^H_{u}$ (stars),
  calculated via the scHA from Eq.~(\ref{Hartree_level}), for
  $\Gamma/\Delta=0.1$ in (b) and $0.4$ in (d). For $\Gamma/\Delta=\infty$
  the Hartree levels are degenerate (not shown).
  The arrow in inset (d) indicates the value of $\e_0$,
    the local minimum of $\e^H_{u}$.}
\label{fig:comp_HF_NRG}
\end{figure}

Since the self-consistent Hartree equation~(\ref{occupation}) may have
more than one solution, a criterion is needed to
pick the correct one. To this end, we note that, for given
$\epsilon$, the system adjusts its local level occupations $n_u$
and  $n_\ell$ such that its total free energy $F_s (n_u, n_\ell)$
is minimized.
Within the scHA approach, $F_s$ can be obtained by
integrating Eq.~(\ref{occupation}), so
that the conditions for $F_s$ to be extremal, $\partial F_s /
\partial n_i = 0$, reproduce Eq.~(\ref{occupation}); we should then
pick that solution of Eq.~(\ref{occupation}) for which the extremum is
a {\it global minimum} of $F_s$.  In the case of {\it two} nearly
degenerate minima, the scHA neglects the possibility of tunneling
between them, and a different approach has to be considered.
Nevertheless we find, somewhat unexpectedly, that in the case of
exactly degenerate minima, e.g. $\Delta=0$, an average over the minima
reproduces the NRG results rather accurately [see
$\Gamma/\Delta=\infty$ curve in Fig.~\ref{fig:comp_HF_NRG}(d)].

For simplicity we implemented this strategy explicitly for
$s=-1$, but not for $s= +1$, since for the
latter ${\cal H}^{\rm eff}_d$  is not diagonal, which
makes the determination of $F_{+1}$ very tedious.
For $s= -1$,   Eq.~(\ref{occupation}) yields the condition
$n_{i}(\e)=\half-\frac{1}{\pi}\arctan\l\{\l(\e_i+Un_{\bari}\r)/\Gamma_i
\r\} \;$,
 and the corresponding free energy has the form
\begin{eqnarray} \label{Functional}
F_{-1} &=& Un_{\ell}n_{u} + \sum_{i = u,\ell} \left[ \e_{i}n_{i} -
{\Gamma_i \over \pi} \log\l(\sin \pi n_i  \r) \right]. \quad
\phantom{.}
\end{eqnarray}
Fig.~\ref{fig:comp_HF_NRG} compare NRG with corresponding scHA results
for $n_i (\epsilon)$.  For $s = -1$, we minimized $F_{-1}$
[Eq.~(\ref{Functional})] and find remarkably good agreement between
NRG and scHA.  For $s = +1$, for which we did not determine $F_{+1}$,
we show the results of a ``naive scHA'', obtained by simply plotting a
numerical solution of Eq.~(\ref{occupation}) and ``hoping'' (without
checking) that it is the correct one. Clearly, the results so obtained
cannot be trusted on their own merit; we present the naive scHA
results nevertheless, to illustrate precisely this point: indeed, in
Fig.~\ref{fig:comp_HF_NRG}(a,c) [for $\Gamma/\Delta>1$] they do
\emph{not} agree well with NRG results.

Fig.~\ref{fig:comp_HF_NRG}(c) and (d) include a special
situation, namely $\gamma=1$ and $\Delta=0$, for which both
$\e_{\ell}=\e_u$ and $\e_{\ell}^{H}=\e_u^H$.  This causes a
sudden jump for $s=+1$ in $n_u$,
but none for $s=-1$ [cf. dashed lines in Fig.~\ref{fig:comp_HF_NRG}(c)
and \ref{fig:comp_HF_NRG}(d), respectively].
To understand why, note that for $\Delta=0$ and  $s=+1$  the odd
local combination $\l(d_{u}-d_{\ell}\r)$ decouples from the leads;
thus, its width is zero and hence its occupation
increases abruptly when its energy drops below zero.  On the other
hand, for $s=-1$ the occupation increases gradually, since the
width of the odd combination is comparable to that of the even one
$\l(d_{u}+d_{\ell}\r)$. A similar argument explains why for
small but non-zero $\Delta/\Gamma$ (odd level almost decoupled for
$s=+1$) as in Fig.~\ref{fig:varyGamma}, curves $3$ and $3'$, the
charge oscillations are still observable for $s=+1$ but not for
$s=-1$.

\section{Unequal coupling amplitudes leading to Occupation inversion}

In addition to providing a simple way to compute physical
quantities, the scHA, and in particular the concept of Hartree
levels [see insets of Fig.~\ref{fig:comp_HF_NRG}(b,d)],  may be used
to qualitatively understand how the physics of the charge
oscillations depends on the various system parameters.  Suppose that
both Hartree levels are swept downwards, starting from $\e$ well
above the Fermi level. When the lower level comes within
$\Gamma_{\ell}$ of the Fermi level, it begins to fill up and the
upper Hartree level $\e^H_u$ is pushed up by $U$, causing a charge
oscillation in $n_u$.  The latter will be stronger the larger $n_u$
was before the oscillation, i.e., the larger the width ($\Gamma_u$)
of the upper level, and the lower the value $\e^H_u(\e_0)$ of the
upper Hartree level at its local minimum, say $\e_0$ (cf.
Fig.~\ref{fig:comp_HF_NRG}).  Indeed, if $\e^H_u(\e_0)\leq\Gamma_u$,
then the upper level achieves a rather significant occupation before
full occupation of the lower level (and corresponding emptying of
the upper one), implying an increase in the amplitude of the
corresponding  charge oscillation.  Moreover, since $\e_u^H(\e_0)$
is also the lower the more suddenly the lower level gets filled, a
smaller $\Gamma_{\ell}$ also strengthens the charge oscillations.
Thus, strong charge oscillations can be obtained quite generally by
allowing $\gamma\equiv\Gamma_{u}/\Gamma_{\ell}\neq1$.  The above
argument implies that $n_u$-oscillations are enhanced for
$\gamma\gg1$; by an analogous argument, with $\ell \leftrightarrow
u$, $n_{\ell}$- oscillations are strengthened for $\gamma\ll 1$.

\begin{figure}[h]
\centering \includegraphics[width=1.0\columnwidth]{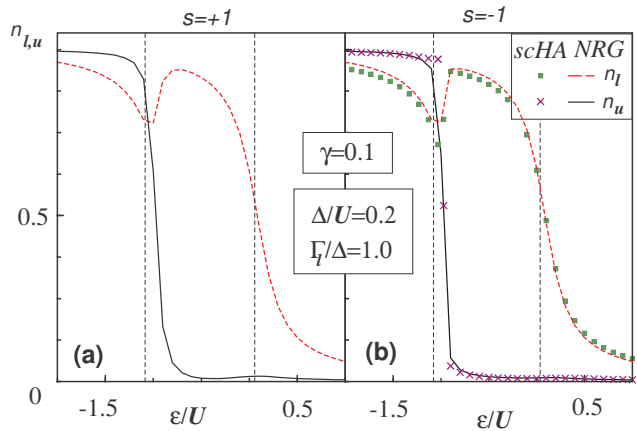}
\caption{Allowing $\gamma \equiv \Gamma_u / \Gamma_{\ell}$ to be
$\neq1$ results in an {\it inversion} of the occupation for a
certain range of $\e$. The scHA results ($n_u$: crosses, $n_\ell$:
boxes) for $s = -1$ (where $F_{-1}$ is known)
 agree well with NRG results ($n_u$: solid, $n_\ell$: dashed lines).
For $\gamma<1$ the Hartree levels cross near the left CB peak,
implying an occupation inversion below the corresponding crossings.
} \label{fig:asym_coupling-1}
\end{figure}

\begin{figure}[h]
 \centering \includegraphics[width=1.0\columnwidth]{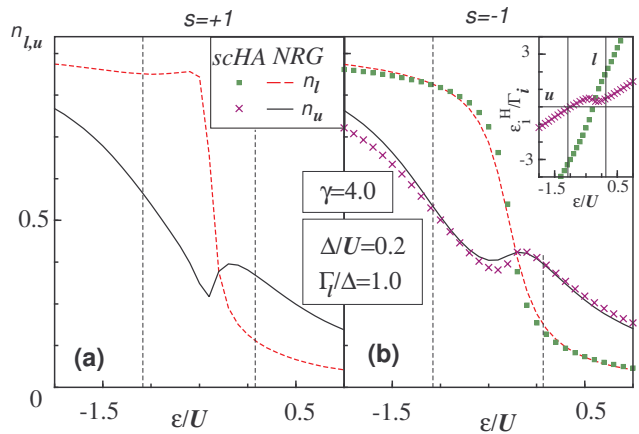}
\caption{
Same as Fig.~\ref{fig:asym_coupling-1}, but for $\gamma > 1$. Now
 the Hartree levels cross near the right CB peak (see
inset (b) for $\gamma=4$), implying an occupation inversion above
the corresponding crossings. Notice the differences in the shape of
the occupation curves between the $s=1$ case and $s=-1$ case. }
\label{fig:asym_coupling-2}
\end{figure}

\begin{figure}[h]
\centering \includegraphics[width=1.0\columnwidth]{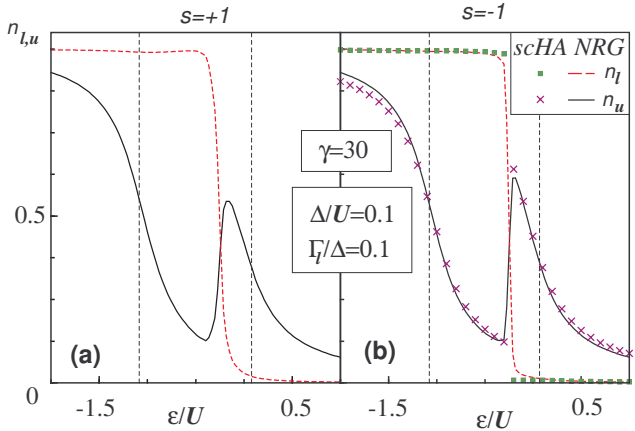}
\caption{(a) NRG results for the occupations with $s=1$ and $\gamma
=30$. (b) A comparison between the scHA results ($n_u$ crosses,
$n_\ell$ boxes) and for the NRG results ($n_u$ solid, $n_\ell$
dashed lines), for $s = -1$ (where $F_{-1}$ is known) and
$\gamma=30$.  The agreement between the NRG and the Hartree
approximation is remarkably good and does not depend on $\gamma$}
\label{fig:asym_coupling-3}
\end{figure}

 It appears that the oscillations are so strong that when $\gamma \neq 1$,
   the lower and the upper Hartree levels might actually cross each other (see inset
Fig.~\ref{fig:asym_coupling-2}), leading to an inverted occupation
[Figs.~\ref{fig:asym_coupling-1},\ref{fig:asym_coupling-2} and
\ref{fig:asym_coupling-3}].

Since the bare energy levels are separated by the level spacing
$\Delta$, the conditions $Un \gtrsim \Delta$ and
$\max\l(\Gamma_{\ell},\Gamma_u\r) \gtrsim \Delta$ must be met to
achieve such an occupation inversion.
Figs.~\ref{fig:asym_coupling-1} to~\ref{fig:asym_coupling-3} show
how the asymmetry ($\gamma\neq1$) of the couplings affects the
occupation of level $\ell$ and $u$ (dashed and solid lines) both for
$s=\pm 1$, leading to an inversion of the occupation within a
certain range of~$\e$.

Our discussion of occupation inversion generalizes a recent related
study by Silvestrov and Imry~\cite{Silvestrov}.  In an attempt to
understand the origin of repeated and abrupt phase lapses observed in
the transmission phase of a multilevel QD~\cite{Schuster}, they
studied a multilevel model consisting of one level (say $u$ with
coupling $\Gamma_u$) strongly coupled to the leads and at least one
additional weakly coupled level (say $\ell$ with coupling
$\Gamma_{\ell}$).  Ref.~\onlinecite{Silvestrov}(a) considered the
limiting case $\Gamma_u \gg \Delta \gg \Gamma_{\ell}\rightarrow 0$
(Ref.~\onlinecite{Silvestrov}(b) also studied finite but small
$\Gamma_{\ell}$ and spin), and compared the energies of the
configuration $(n_{u},n_{\ell})=(1,0)$ to that of
$(n_{u},n_{\ell})=(0,1)$ in second order perturbation theory in the
tunneling. Their results indicate that the system is able to sustain
an occupation inversion until
$\e_u=-U/\l[\exp\l(2\pi\Delta/\Gamma_u\r)+1\r]\approx -U/2$.  Although
we approach this problem from a different angle, i.e., we either solve
it exactly by NRG or first solve the tunnel-coupling exactly and then
treat the interaction self-consistently, the inversion range found in
Ref.~\onlinecite{Silvestrov} coincides with the
results~\cite{inversion} of this article.

Our analysis indicates that the example of Ref.~\onlinecite{Silvestrov} for
occupation inversion is a very special case of a  much
more general phenomenon whose
strength depends on $\gamma$: as $\gamma$ is increased from 1
 (where no occupation inversion occurs), (i) the range of
gate voltages in which inversion occurs
increases, with
the inversion point moving towards the middle of the CB valley;
and (ii) the maximal value reached by $n_u$ right before the
inversion increases gradually towards $1$, i.e., the effect
becomes more pronounced.

 \section{Summary}
\label{sec:summary}

In this paper, we have studied the gate voltage dependence of
the occupation of a  spinless two-level Anderson model for the
generic case of a relative sign $s$ in the tunneling amplitude. We
found a non-monotonic behavior in the occupation of the local
levels, due to charging effects between electrons within the QD, and
explained this effect in the framework of a self-consistent Hartree
approximation. Remarkably, the occupations of the upper and lower levels can
even be inverted if the level-to-lead couplings are sufficiently
asymmetric. Even though we focused on $T=0$ throughout this paper we expect
the calculated behavior to persist as long as $T\lesssim\min\{\Gamma,\Delta,U\}$.

\begin{figure}[h]
 \centering \includegraphics[width=1.0\columnwidth]{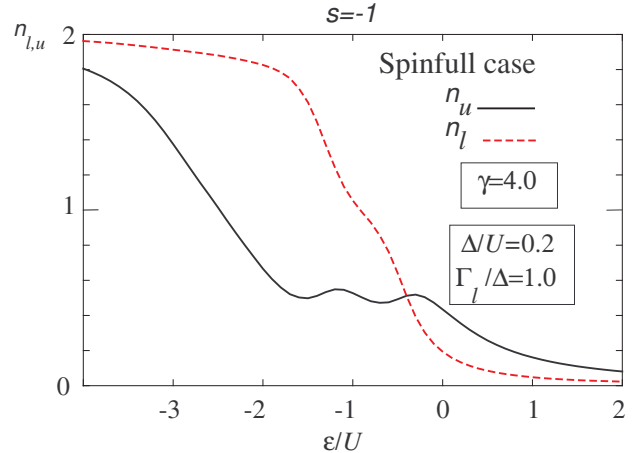}
\caption{ NRG results in the presence of spin (with the relative
sign $s=-1$). The parameters are identical to those in
Fig.~\protect{\ref{fig:asym_coupling-2}}. The occupation crossing
persists in the spinful case. The additional oscillations in the
occupations are observed due to the possibility to put two electrons
of opposite spin with in each level.} \label{fig:NRGspin}
\end{figure}

The inclusion of spin in the 2LAM, though making the
problem more complex, does not change the qualitative
results presented above in a large region of parameter space.
To illustrate this, we show in Fig.~\ref{fig:NRGspin} the total
occupation of the upper and lower levels obtained by NRG for the
spinful model.  The level crossing persists and additional oscillations
are observed due to the possibility to put two electrons of opposite
spin within each level.

Our predictions of the non-monotonicity of the charging of a 2LAM are
experimentally relevant (in spirit, if not in detail) for any
quantum dot system containing orbital levels that are ``nearly
degenerate'', in the sense that their spacing is smaller than the
level widths. One way of realizing the specific models studied here
would be to use two capacitively coupled quantum dots, each with
large level spacings, associating their topmost not-fully-occupied
levels with $\epsilon_u$ and $\epsilon_\ell$, and using a large
magnetic field to lift the Zeeman degeneracy of each. The way in
which the charges on these dots evolve with gate voltage, i.e., the
evolution of $n_i (\epsilon)$, could then be measured experimentally
using QPC's serving as extremely sensitive charge sensors, see e.g.
Fig.~1 (a) of Ref.~\onlinecite{Sprinzak}.
In addition, we expect that the occupation of the levels will be reflected in
other properties of the system such as the transmission phase.
\cite{Schuster} We leave that, however, for future studies.

\section{acknoweldgements}

We thank R. Berkovitz, L. Borda, Y. Gefen, U. Hartmann, M. Heiblum,
F. Hekking, J. K\"onig, F. von Oppen, and G. Zarand  for clarifying
discussions. Financial support from SFB 631 of the DFG, CeNS,
Minerva, DIP c-7.1 and ISF 845/04 is gratefully acknowledged. While
performing this work, we became aware of parallel work~\cite{Gefen}
that is closely related to ours.


\begin{thebibliography}{10}
\bibitem{ADeltaU} In GaAs-AlGaAs, a QD of area
  $A\approx 0.1 \mu \textrm{m} \times  0.1\mu \textrm{m}$ has
  $\Delta \approx 0.5 \textrm{meV} \approx 6 \textrm{K}$
  and $U \approx 1.5 \textrm{meV}$.
\bibitem{validity} For $V_{uR}=V_{\ell L}=0$, our model
   includes the case of two dots (levels $u$, $\ell$) separately
   coupled to two leads $(L,R)$.
\bibitem{phsymmetry} For $\Gamma_{\ell}=\Gamma_u$ (and both
  $s=\pm1$), the system is particle-hole-symmetric under $c_{k}
  \ra{\tilde c_{-k}^{\dag}}$ and $d_{i}\ra -{\tilde
    d_{\bar i}^{\dag}}$ where $\bari=\ell/u$ if $i=u/\ell$, implying
  $n_{i}\ra (1-{\tilde n_{\bar i}})$.
\end{thebibliography}
\end{document}